\documentclass[journal=jacsat,manuscript=article]{achemso}

\usepackage[version=3]{mhchem} 
\usepackage[T1]{fontenc}       
\usepackage[usenames, dvipsnames]{color}
\usepackage{longtable}
\usepackage{multirow}
\usepackage{makecell}

\author{Duy Khanh Nguyen}
\email{nguyenkhanhphysics2015@gmail.com}
\author{Yu-Tsung Lin}
\author{Shih-Yang Lin}
\affiliation[National Cheng Kung University]
{Department of Physics, National Cheng Kung University, 701 Tainan, Taiwan}

\author{Yu-Huang Chiu}
\affiliation[National Pingtung University]
{Department of Applied Physics, National Pingtung University, 900 Pingtung, Taiwan}
\author{Ngoc Thanh Thuy Tran}
\author{Ming Fa-Lin}
\email{mflin@mail.ncku.edu.tw}
\phone{+886-6-275-7575}
\fax{886-6-274-7995}
\affiliation[National Cheng Kung University]
{Department of Physics, National Cheng Kung University, 701 Tainan, Taiwan}

\title{Fluorination-Enriched Electronic and Magnetic Properties in Graphene Nanoribbons}

\begin{document}


\begin{abstract}

 The feature-rich electronic and magnetic properties of fluorine-doped graphene nanoribbons are investigated by the first-principles calculations. They arise from the cooperative or competitive relations among the significant chemical bonds, finite-size quantum confinement and edge structure. There exist C-C, C-F, and F-F bonds with the multi-orbital hybridizations. Fluorine adatoms can create the p-type metals or the concentration- and distribution-dependent semiconductors, depending on whether the \( \pi  \) bonding is seriously suppressed by the top-site chemical bonding. Furthermore, five kinds of spin-dependent electronic and magnetic properties cover the non-magnetic and ferromagnetic metals, the non-magnetic semiconductors, and the anti-ferromagnetic semiconductors with/without the spin splitting. The diverse essential properties are clearly revealed in the spatial charge distribution, the spin density, and the orbital-projected density of states.  

 \noindent \textit{Keywords}: Multi-orbital hybridizations, fluorination, top-site doping, buckled structures, strong electron affinity
\end{abstract}

\section{Introduction}
\par\noindent

A new scientific era has been opened since the discovery of graphene \cite{1}. This 2D system exhibits a lot of novel and unusual electronic properties \cite{2}. However, there exist few obstacles in the potential applications of graphene-based materials \cite{3}. To overcome the gapless feature, the direct method is to create one-dimensional (1D) strips of graphene, usually referred to as graphene nanoribbons \cite{4}. GNRs are one of the main-stream nanomaterials, mainly owing to the complex relations among honeycomb lattice, one-atom thickness, finite-size quantum confinement and edge structure. Each GNR could be regarded as a finite-width graphene strip or an unzip carbon nanotube \cite{5}. Up to now, GNRs have been successfully synthesized by the various experimental methods including both top-down and bottom-up schemes \cite{6}. From the geometric point of view, graphene cutting seems to be the simplest and intuitive method, in which the available routes cover lithographic patterning \cite{7} and etching of graphene \cite{8}, sonochemical breaking \cite{9}, metal-catalyzed cutting of graphene \cite{10}, and oxidation cutting of graphene. Another approach is to unzip carbon nanotube using metal nano-clusters as scalpels \cite{11}, and a wet chemical method based on acid reactions \cite{12}. The chemical vapor deposition is utilized to massively produce GNRs to meet the essential requirement of semiconductor industry \cite{13}. GNRs are expected to have high potential applications in nano-electronic \cite{14} and spintronic \cite{15} devices, gas sensor \cite{16}, and nanocomposites \cite{17}.

Interestingly, electronic properties of GNRs can be easily modulated by chemical doping \cite{18}, mechanical strain \cite{19,20}, layer number \cite{21}, curved surface \cite{22}, edge-passivation \cite{23,24}, stacking configuration \cite{25}; electric \cite{26,27} and magnetic \cite{28,29} fields. Among these modulations, the chemical modification on ribbon surface is the most effective one in creating the dramatic changes between the semiconducting and metallic behaviors (the non-magnetic and magnetic configurations). The previous theoretical studies clearly show the geometry- and doping-enriched electronic and magnetic properties. Two typical achiral GNRs, armchair and zigzag ones (AGNRs and ZGNRs), present the width-dependent energy gaps \cite{30,31}, and the latter possess the anti-ferromagnetic spin configuration across the ribbon center \cite{32}. The chemical dopings of transition metal Co/Ni adatoms will induce the metallic band structures with free conduction electrons \cite{33}, in which the spin-split energy bands correspond to the ferromagnetic configuration.  However, alkali adatoms can create the non-magnetic metals in AGNRs and the ferromagnetic ones in ZGNRs under specific distributions \cite{34}. The ligand-protected aluminum clusters adsorbed AGNRs lead to the semiconducting or metallic band structures, depending on their kinds \cite{35}.  As for molecule adsorptions, (CO, NO, NO\(_2\), O\(_2\), N\(_2\), CO\(_2\)) do not change the semiconducting behavior \cite{36}, while NH\(_3\) presents the n-type doping. On the experimental side, the adsorption of hydrogen molecules on the Pd-functionalized multi-layer GNRs are successfully obtained \cite{37}, and tin oxide nanoparticles are also synthesized on GNRs to form a composite material \cite{38}. These results indicate that surface chemical adsorption may serve as a tool for controlling the electronic properties of GNRs \cite{39}. However, a systematic theoretical study on the halogen-adsorbed GNRs is absent up to now. Fluorine adatoms have very strong electron affinity; they are thus expected to present the significant chemical bondings with carbon atoms and greatly diversify the essential properties.

This work is focused on the essential geometric, electronic and magnetic properties of fluorine-doped GNRs. They are explored in detail by using the density functional theory. The bond lengths, positions of adatoms, ground state energies, energy bands, spatial charge distributions, free carrier densities, spin densities, magnetic moments, and density of states (DOS) are evaluated using the first-principles calculations. The dependence on concentration, distribution of adatoms, and edge structure is fully included in the calculations. By the detailed analyses, the critical orbital hybridizations in chemical bonds are identified from atom-dominated energy bands, the spatial charge distribution, and the orbital-projected DOSs. The current study shows that they are responsible for the diverse electronic and magnetic properties, covering the ferromagnetic and non-magnetic metals, the non-magnetic semiconductors, and the anti-ferromagnetic semiconductors with/without spin splitting. Furthermore, the feature-rich band structures are reflected in a lot of prominent peaks in DOSs. The predicted optimal geometries, energy bands and DOSs could be verified by scanning tunneling microscopy (STM) \cite{40}, angle-resolved photoemission spectroscopy (ARPES) \cite{41} and scanning tunneling spectroscopy (STS) \cite{42}, respectively. 
\section{Computational methods}
\par\noindent

The essential properties of F-doped GNRs are investigated by using the Vienna ab initio simulation package \cite{43} within the spin-polarized density functional theory. The exchange and correlation energies, which come from the many-particle Coulomb interactions, are evaluated from the Perdew-Burke-Ernzerhof functional \cite{44} under the generalized gradient approximation. Furthermore, the projector-augmented wave pseudopotentials can characterize the electron-ion interactions \cite{45}. Plane waves, with an maximum energy cutoff of \(400\)
 eV, are utilized to calculate wave function and state energies. The 1D periodic boundary condition is along 
 \(
 \hat x \), and the vacuum spacing associated with  \(
  \hat y
  \)  and  \(
   \hat z
   \)  is larger than \( 15 \)
  \AA  \, to avoid the interactions between two neighboring nanoribbons. The Brillouin zone is sampled by  \( 15 \times 1 \times 1 \)
   and \( 100 \times 1 \times 1 \) k point meshes within the Monkhorst-Pack scheme for geometric optimizations and further calculations on electronic structures, respectively. The convergence for energy is set to be 
   \(10^{ - 5}   \)
   eV between two simulation steps, and the maximum Hellmann-Feynman force acting on each atom is less than \(  0.01  \)
    eV/\AA \, during the ionic relaxations.

\section{Results and discussion}
\par\noindent

	The geometric, electronic and magnetic properties of fluorine-adsorbed GNRs are investigated for various distributions and concentrations of adatoms in zigzag and armchair systems. The widths of AGNR and ZGNR, as shown in Figs. 1(a) and 1(b), are characterized by the number of dimers lines and zigzag lines (N\(_A \)
 and N\(_Z \)) along 
 \(\hat y\), respectively, in which the periodical lengths in a unit cell along \(\hat x\)  are 
 \(3b\)
   and \( 2\sqrt 3 b \) (\(b\)  the C-C bond length). In general, the double-side adsorptions have the lower ground state energies    \(E_0\)'s ,  compared to the single-side cases (Table 1). The optimal adatom position is situated at the top site, regardless of any doping cases. Fluorination can induce the buckled GNR structure. Carbon atoms nearest to F deviate from the graphene plane, being sensitive to distributions and concentrations. For single adatom adsorption, the carbon heights are, respectively, \(0.053\) \AA \, and \(0.133\) \AA \,  at center and edge of GNR. They obviously grow with F-concentration, e.g., \(0.205\) \AA \, for 
   \( 100\%  \)  adsorption. F adatoms are very close to C, in which the shortest and longest F-C bond lengths (\( 1.395\) \AA \, and \(1.547\) \AA \,), respectively, correspond to the highest concentration and single adatom near ribbon center. Moreover, the nearest C-C bond lengths are lengthened in the range of 
   \(0.06 - 0.12 \) \AA \,, compared with those of pristine GNRs. This indicates the \(\sigma\)-bonding changes due to the strong fluorination. The critical F-C chemical bondings, being responsible for the featured geometric structures, are expected to dominate the other essential properties.

                    \begin{table}[htb]
                    \caption{Ground state energy, magnetic moment and magnetism, energy gap, free holes in a unit cell, and geometric parameters for N\(_A=12\) armchair and N\(_Z=8\) zigzag GNRs under single- and double-side fluorinations. NM, FM and AFM, respectively, correspond to non-magnetism, ferro-magnetism and anti-ferro-magnetism.}
                         \label{t1}
                           \begin{center}
                                                
                           \begin{tabular}{ |l|l|l|l|l|l|l|l|l|}
                                \hline
                                                GNRs & \makecell{Adsorption\\ configurations }  & \(E_0\) (eV)
                                                 & \makecell{Magnetic\\ moment \\(\(\mu _B \) )/\\magnetism } & \makecell{ \( E_g^{d(i)} \) \\(eV)/ \\Metal}  & \makecell{Num-\\ber\\of\\holes} & \makecell{F-C\\ (\AA)}  & \makecell{C \\height\\(\AA)}  & \makecell{Nearest\\C-C\\(\AA)}  \\ 
                                                \hline
                                                \multirow{21}{4em}{AGNR\\N\(_A=12\)} & Pristine & -234.7419 & 0/NM &  \( E_g^{d}=0.6 \)& 0 & 0 & 0 & 1.428 \\ 
                                                 & \((13)_s\) & -237.0912 & 0/NM & M & 1 & 1.547 & 0.053 & 1.483 \\ 
                                                 & \((1)_s\) & -237.4201 & 0.47/FM & M & 1 & 1.471 & 0.133 & 1.489 \\ 
                                                 & \((6, 21)_s\) & -239.4138 & 0/NM & M & 2 & 1.545 & 0.055 & 1.491 \\ 
                                                 & \((6,\textcolor{red}{21})_d\) & -239.4261 & 0/NM & M & 2 & 1.546 & 0.054 & 1.491 \\ 
                                                 & \((1, 23)_s\) & -240.4703 & 0.76/FM & M & 1 & 1.468 & 0.132 & 1.488 \\ 
                                                 & \((1,\textcolor{red}{23})_d\) & -240.4839 & 0.76/FM & M & 1 & 1.469 & 0.131 & 1.488 \\ 
                                                 & \((1,\textcolor{red}{6})_d\) & -241.0514 & 0/NM & \( E_g^{i}=0.64 \) & 0 & 1.443 & 0.161 & 1.493 \\ 
                                                 & \((6,\textcolor{red}{9},14,\textcolor{red}{17})_d\) & -245.3591 & 0/NM & M & 2 & 1.503 & 0.097 & 1.497 \\ 
                                                 & \((2,\textcolor{red}{9},18,\textcolor{red}{23})_d\) & -245.8006 & 0.56/FM & M & 1 & 1.454 & 0.15 & 1.491 \\ 
                                                 & \((1,\textcolor{red}{2},\textcolor{red}{23},24)_d\) & -250.2258 & 0/NM & \( E_g^{i}=0.98 \) & 0 & 1.43 & 0.17 & 1.496 \\ 
                                                 & (C:\textcolor{red}{F}=24:\textcolor{red}{6})\(_d\) & -250.4901 & 0/NM & M & 2 & 1.532 & 0.068 & 1.497 \\ 
                                                 & (C:\textcolor{red}{F}=24:\textcolor{red}{6})\(_d\)& -251.897 & 0.57/FM & M & 1 & 1.443 & 0.161 & 1.493 \\ 
                                                 & (C:\textcolor{red}{F}=24:\textcolor{red}{6})\(_d\) & -255.5256 & 0/NM & \( E_g^{i}=0.85 \) & 0 & 1.414 & 0.186 & 1.506 \\ 
                                                 & (C:\textcolor{red}{F}=24:\textcolor{red}{8})\(_d\)  & -258.6336 & 0/NM & M & 2 & 1.446 & 0.154 & 1.503 \\ 
                                                 & (C:\textcolor{red}{F}=24:\textcolor{red}{8})\(_d\)  & -257.9218 & 0.59/FM & M & 1 & 1.473 & 0.127 & 1.498 \\ 
                                                 & (C:\textcolor{red}{F}=24:\textcolor{red}{8})\(_d\) & -261.6414 & 0/NM & \( E_g^{i}=0.12 \) & 0 & 1.416 & 0.184 & 1.506 \\ 
                                                 & (C:\textcolor{red}{F}=24:\textcolor{red}{10})\(_d\) & -268.196 & 0/NM & \( E_g^{i}=0.56 \) & 0 & 1.416 & 0.184 & 1.507 \\ 
                                                 & (C:\textcolor{red}{F}=24:\textcolor{red}{14})\(_d\) & -283.013 & 0/NM & \( E_g^{i}=2.25 \) & 0 & 1.413 & 0.187 & 1.53 \\ 
                                                 & (C:\textcolor{red}{F}=24:\textcolor{red}{20})\(_d\) & -303.9926 & 0/NM & \( E_g^{i}=2.69 \) & 0 & 1.408 & 0.192 & 1.505 \\ 
                                                 & (C:\textcolor{red}{F}=24:\textcolor{red}{24})\(_d\) & -319.2921 & 0/NM & \( E_g^{d}=3.2 \) & 0 & 1.395 & 0.205 & 1.544 \\ 
                                                \hline
                                                \multirow{6}{4em}{ZGNR\\N\(_Z=8\)} & Pristine & -308.0119 & 0/AFM & \( E_g^{d}=0.46 \) & 0 & 0 & 0 & 1.428 \\ 
                                                 & \((3)_s\) & -311.9155 & 0.42/FM & M & 1 & 1.462 & 0.138 & 1.485 \\ 
                                                 & \((13)_s\)  & -310.4937 & 0.4/FM & M & 1 & 1.55 & 0.05 & 1.481 \\ 
                                                 & \((3,\textcolor{red}{14})_d\) & -314.6181 & 0.37/FM & M & 1 & 1.45 & 0.15 & 1.488 \\ 
                                                 & \((3,\textcolor{red}{6})_d\) & -314.8808 & 0.37/FM & M & 1 & 1.432 & 0.168 & 1.492 \\ 
                                                 & \((11,\textcolor{red}{14})_d\) & -313.5967 & 0/AFM & \( E_g^{i}=0.2 \) & 0 & 1.501 & 0.099 & 1.494 \\ 
                                                 & \((19,\textcolor{red}{22})_d\) & -313.3396 & 0/AFM & \( E_g^{i}=0.2 \) & 0 & 1.504 & 0.096 & 1.486 \\ 
                                                 & \((3, 30)_s\) & -315.9246 & 0/NM & \( E_g^{d}=0.46 \) & 0 & 1.454 & 0.146 & 1.487 \\ 
                                                 & \((3,\textcolor{red}{30})_d\) & -315.9294 & 0/NM & \( E_g^{d}=0.46 \) & 0 & 1.455 & 0.145 & 1.487 \\ 
                                                 & (C:\textcolor{red}{F}=32:\textcolor{red}{32})\(_d\) & -420.1127 & 0/NM & \( E_g^{d}=2.97 \) & 0 & 1.381 & 0.219 & 1.515 \\ 
                                                \hline
                                                \end{tabular}
                                                 \end{center}
                                                 \end{table}

    \begin{figure}[htb]
                \includegraphics[width=8cm, height=20cm]{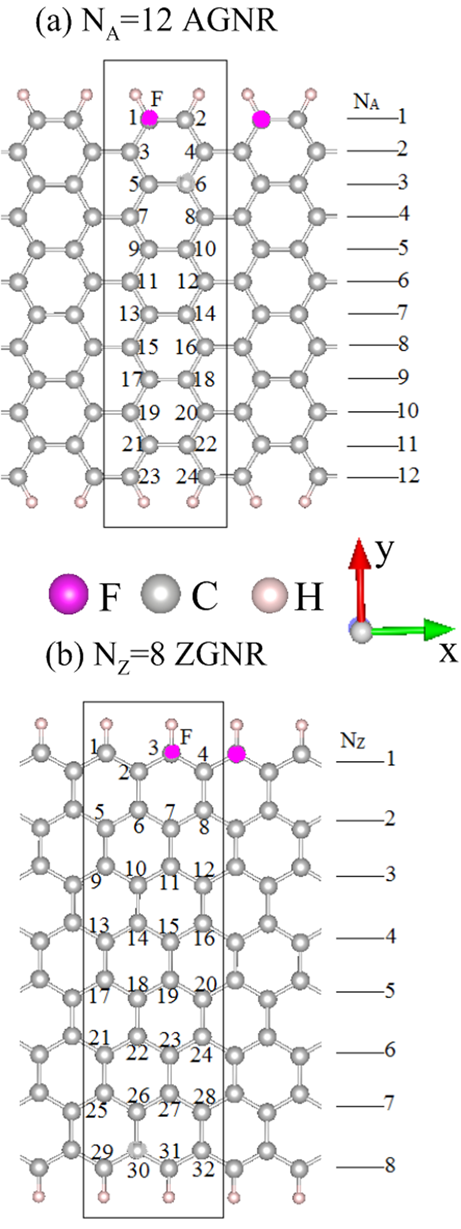}
                \caption{Geometric structures of F-adsorbed GNRs for (a) N\(_A=12\) armchair and (b) N\(_Z=8\) zigzag systems. The black rectangles represent unit cells. The lattice constants are, respectively, \(a=3b\)  and 
                \(
                a = 2\sqrt 3 b
                \)
                  for armchair and zigzag GNRs. Numbers on the top of carbons denote the positions of adatoms.}
                \label{fgr:1}
              \end{figure}
   
   GNRs possess the unusual 1D band structures because of honeycomb symmetry, quantum confinement and edge structure.  Pristine AGNRs present a lot of 1D energy bands, as shown in Fig. 2(a) for the N\(_A=12\) system. The occupied valence bands are asymmetric to the unoccupied conduction bands about the Fermi level 
   \( E_F \), in which a direct energy gap of   \( E_g^d  = 0.6  \)  eV at the \(\Gamma \) point arises from the finite-size effect. The electronic states of  \( E^{c,v}  \le 2  \)
    eV and the deeper ones are, respectively, dominated by the \(\pi\) bondings of parallel  2p\(_z  \)
      orbitals, and the \(\sigma\) bondings of (2p\(_x\),2p\(_y\)) orbitals (indicated from the orbital-projected DOSs in Fig. 6(a)).  Most of energy bands belong to parabolic dispersions, while few of them have partially flat ones within a certain range of \(k_x \) (e.g., \( E^v  =  - 2.1\) eV and \( - 4.7\) eV). All the energy dispersions depend on wave vector monotonously except for the subband anti-crossings.  The band-edge states, which occur at \(k_x=0,1 \) (in unit of
      \( \pi /3b \) ), and others related to subband anti-crossings, will create the van Hove singularities in DOSs.

      The semiconducting band structures are dramatically changed by the strong fluorination. For the single-adatom adsorption, the F-doped AGNRs exhibit the metallic energy bands, as clearly indicated from \((13)_s\) in Figs. 2(b) and \((1)_s\) in Fig. 2(c). The Fermi level is shifted to the \(\pi\)-electronic valence states; that is, \(E_F\) presents a red shift.  There exist free holes in the unoccupied valence states between two Fermi momenta (
      \(\pm k_F \)  related to two valence bands intersecting with \(E_F\)). Electrons are transferred from carbon atoms to fluorine adatoms. The low-lying energy bands mainly arise from the \(\pi\) bondings of carbon atoms, being independent of adatom positions. Moreover, the (F,C)-co-dominated energy bands, accompanied with the \(\sigma\) bands, appear at   \( E^v  \le  - 2.5\)  eV. They have the weak energy dispersions or the narrow band widths. The rich features of energy bands further illustrate the critical F-C bondings, leading to the significant modifications of \(\pi\) and \(\sigma\) bands (\(\pi\) and \(\sigma\) bondings).
      
      The main features of electronic structures are very sensitive to the variations in the concentration, relative position, single- or double-side adsorptions, and edge structure.  With the gradual increase of concentration, the Fermi-momentum states are drastically changed, as shown in Figs. 2(d)-2(h) for two-F adsorptions. The total free hole density, the summation of Fermi momenta in partially unoccupied \(\pi\)-electronic valence bands, will become higher [\((6,21)_s\) in Fig. 2(d) and \((6,\textcolor{red}{21})_d\) in Fig. 2(e); Table 1] or keep the same [\((1,23)_s\) in Fig. 2(g)  and \((1,\textcolor{red}{23})_d\) in Fig. 2(h)], compared to that of the single adatom (Figs. 2(b) and 2(c)).  Moreover, the metallic band structure might be thoroughly changed into the semiconducting one for a very close adatom distance, e.g., an indirect gap of  \(E_g^i  = 0.64  \)
        eV for the \((1,\textcolor{red}{6})_d\) adsorption in Fig. 2(f). This suggests the termination of the extended \(\pi\) bonding in AGNRs. Specifically, all the F-doped AGNRs belong to larger-gap semiconductors under high adatom concentrations (Figs. 2(j)-2(l)), in which the critical concentration is estimated to be about \(10/24\) in the N\(_A=12 \) system (Fig. 2(j)). There are more F-dependent valence bands. Such energy bands determine the magnitude of energy gap, and \( E_g \)
        also depends on the \(\sigma\) bonding of carbon atoms (DOSs in Figs. 6(c)-6(d)). A large gap of \(E_g^i  = 3.2 \)  eV is revealed in the highest adatom concentration (
        \(100\% \) adsorption in Fig. 2(l)). In addition, the single- and double-side adsorptions present the almost identical low-lying energy bands and spin configurations (e.g., Figs. 2(g) and 2(h); Figs. 4(b) and 4(c)), when the \((x,y)\) projection and the adatom concentration keep the same.
      
      \begin{figure}[htb]
                      \includegraphics[width=8cm, height=20cm]{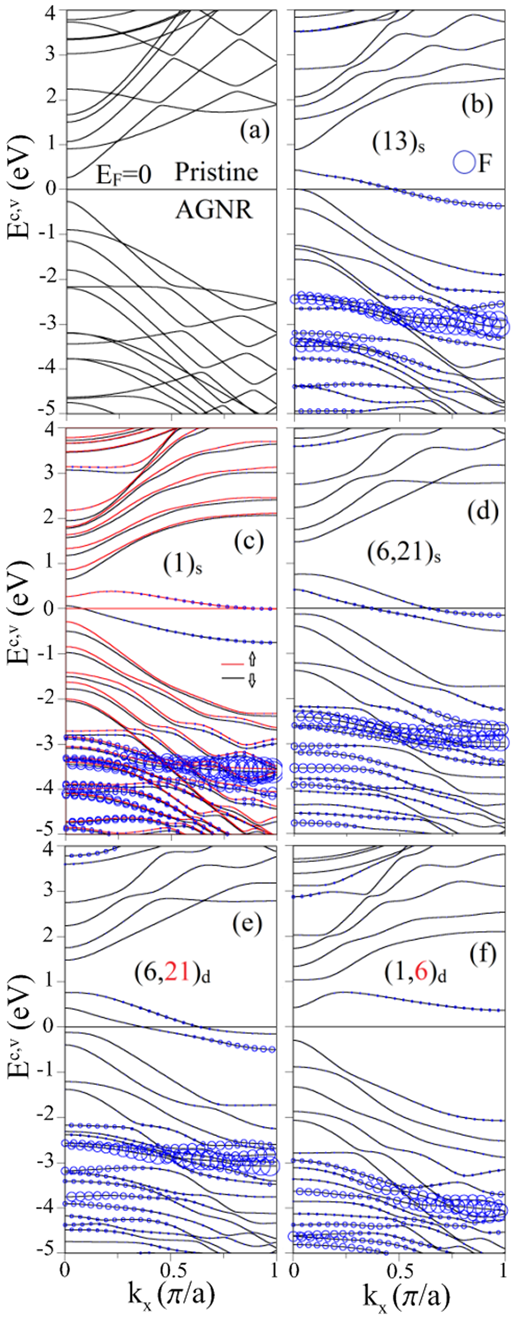}
                         \label{fgr:2}
                   \end{figure}
                   
                   \begin{figure}[htb]
                                  \includegraphics[width=8cm, height=20cm]{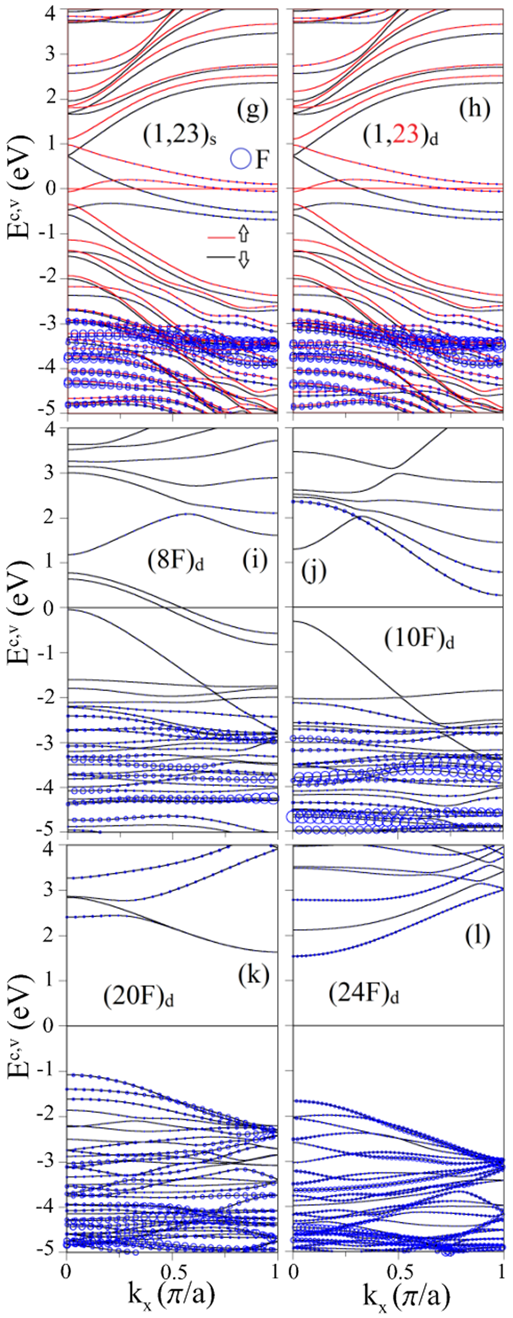}
                                     \caption{Band structures of N\(_A=12\) AGNR for (a) pristine, (b) \((13)_s\), (c) \((1)_s\), (d) \((6, 21)_s\), (e) \((6,\textcolor{red}{21})_d\), (f) \((1,\textcolor{red}{6})_d\),  (g) \((1,23)_s\), (h) \((1,\textcolor{red}{23})_d\), (i) (8F)\(_d\), (j)
                                     (10F)\(_d\), (k) (20F)\(_d\), \(\&\) (l) (24F)\(_d\); Blue circles represent the contribution of F adatoms. The red and black curves denote the spin-split energy bands. }
                                     \label{fgr:2}
                               \end{figure}

       The linear free hole density deserves a closer examination. It is linearly proportional to the Fermi momentum by the relation \(\lambda  = 2k_F /\pi\)
         for each partially unoccupied valence band. By the detailed calculations and analyses, \(\lambda\)  has no simple relation with concentration except for one-adatom adsorption. A single F in a unit cell can create almost one free hole for any adsorption positions (Table 1), i.e., it attracts one electron from the bonded carbon atom because of rather strong electron affinity. The metallic 4-, 6-, and 8-adatom adsorptions can present two or one free holes in a unit cell, respectively, corresponding to the spin-degenerate and spin-split energy bands in Fig. 2 (non-magnetism and ferro-magnetism). However, only the lower carrier case is revealed in zigzag systems (Figs. 3(b) and 3(f); Table 1).  In addition, when the fully unoccupied \(\pi\)-electronic valence states are taken into account, each F adatom just contributes one hole in a unit cell. The F-doped GNRs are in sharp contrast with the alkali-doped systems \cite{46}. The latter belongs to n-type metals even for the \(100\% \)  adatom concentration. Each alkali adatom generates one conduction electron from the outermost s orbital by means of the significant alkali-C bond, being independent of adatom distributions. This will lead to very high conduction electron density. 
       
       Edge structures play a critical role in the diverse essential properties. There are certain important differences between zigzag and armchair systems in band structures without or with fluorination.  Pristine ZGNRs, as shown in Fig. 3(a), have a pair of partially flat valence and conduction bands nearest to \(E_F\)  at small \(k_x\)'s, corresponding to wave functions localized at the zigzag boundaries \cite{47}. Such bands have the double degeneracy for the spin degree of freedom even if they are closely related to the anti-ferromagnetic configuration across the ribbon center and the ferromagnetic one at the same edge (discussed in Fig. 4(d)). Their energy dispersions become strong at large \(k_x\)'s. The band-edge states, which appear at \( k_x  = 1/2\), determine a direct gap of 
       \(E_g^d  = 0.46\) 
         eV for a N\(_Z=8\) ZGNR. This gap is due to the strong competition between quantum confinement and spin configuration. The partially flat bands near \(k_x=0\) and \(E_F\), with the localized charge distributions, might be changed by fluorination, such as, energy dispersions, energy gap, and state degeneracy in Figs. 3(b)-3(f). When one F adatom is very close to the zigzag edge, the number of edge-localized energy bands is  reduced to half in the presence of spin splitting, as shown for the \((3)_s\) adsorption in Fig. 3(b). Furthermore, such bands intersect with the Fermi level and thus exhibit the metallic behavior (Table 1). They will disappear under two adatoms near both zigzag edges [\((3,30)_s\) in Fig. 3(c) and \((3,\textcolor{red}{30})_d\) in Fig. 3(d)], leading to a direct-gap semiconductor (Table 1). As for the central two-F adsorption, the double of partial flat bands, with spin splitting, come to exist [\((11,\textcolor{red}{14})_d\) in Fig. 3(e)], in which half of them correspond to the edge- or center-localized electronic states. This system is an indirect narrow-gap semiconductor. Specifically, the number of low-lying flat bands keeps the same if two adatoms are, respectively, close to edge and center [\((3,\textcolor{red}{14})_d\) in Fig. 3(f)].  Their spin splitting creates a 1D metal. Also, it should be noticed that all the metallic F-absorbed ZGNRs, being related to the spin-split partially flat bands, present one free hole in a unit cell (Table 1). 
         
          The dependence of energy bands on electron spins is diversified by fluorine doping and edge. There exist five kinds of spin-dependent electronic and magnetic properties. The pristine AGNRs are confinement-induced semiconductors without spin-split energy bands and magnetism (the first kind in Fig. 2(a) and Table 1). The similar features are also revealed in the semiconducting F-doped AGNRs (Figs. 2(f) and 2(j)). However, the metallic systems might present the spin-degenerate energy bands in the absence of magnetism (the second kind in Figs. 2(d) and 2(e)), or the spin splitting with the ferromagnetic configuration (the third kind in Figs. 2(g) and 2(h)). As for ZGNRs, the pristine systems are the anti-ferromagnetic semiconductors without spin splitting (the fourth kind in Fig. 3(a)); furthermore, the F-doped ones present the first kind (Figs. 3(c) and 3(d)), the third kind (Figs. 3(b) and 3(f)), or the fifth kind (the semiconducting behavior with spin splitting under the anti-ferromagnetic configuration; Fig. 3(e)).
          
          \begin{figure}[htb]
                             \includegraphics[width=8cm, height=20cm]{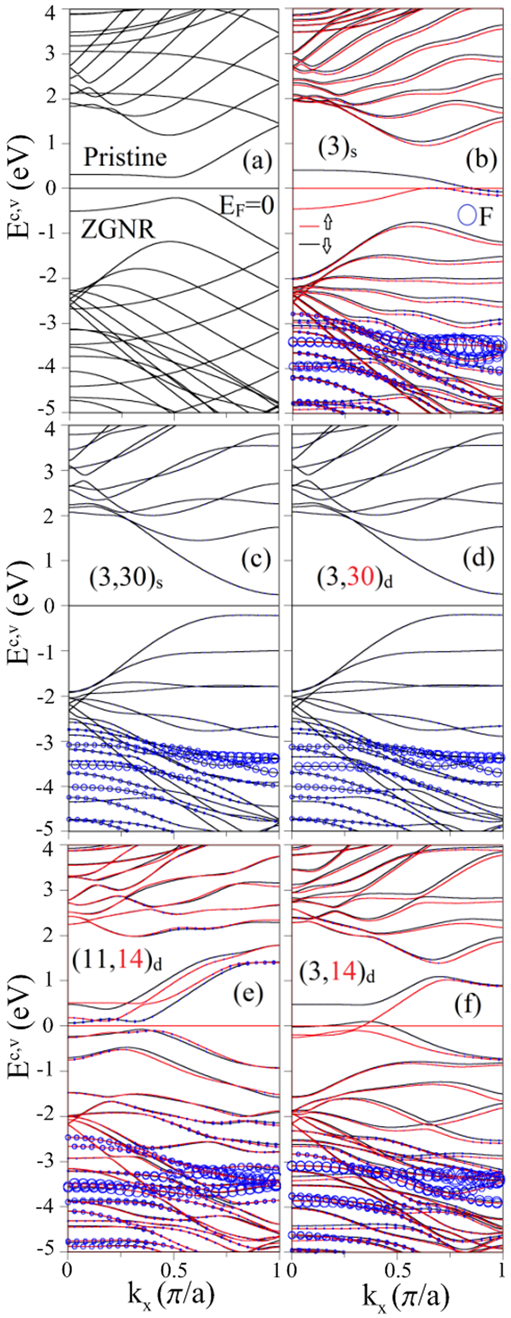}
                              \caption{Band structures of N\(_Z=8\) ZGNR for (a) pristine ZGNR, (b) \((3)_s\), (c) \((3, 30)_s\), (d) \((3,\textcolor{red}{30})_d\), (e) \((11,\textcolor{red}{14})_d\), \(\&\) (f) \((3,\textcolor{red}{14})_d\); Blue circles represent the contribution of F adatoms. The red and black curves denote the spin-split energy bands.}
                             \label{fgr:3}
                             \end{figure}

          The spatial spin densities and magnetic moments could provide more information about the essential properties.  The third, fourth and fifth kinds exhibit the different arrangements (Figs. 4(a)-4(d), Fig.  4(e) and Fig. 4(f)), in which the competition between the spin-up and spin-down  configurations determines the net magnetic moment in a unit cell (Table 1). The spin densities are mostly distributed near the edge structures except for the fifth kind (Fig. 4(f)). When one F is situated at the armchair (zigzag) edge, the ferromagnetic spin-up configuration occurs there (the similar one is created at another one), as shown in Fig. 4(a) (Fig. 4(d)). This leads to a net moment of  \(0.47\) \((0.42)\) \(\mu _B  \). For two-F adsorptions near both boundaries, AGNRs exhibit the enhanced ferromagnetism across the ribbon center with \(0.76\)  \(\mu _B  \) (Figs. 4(b) and 4(c)), while magnetism is absent in ZGNRs (Figs 3(c) and 3(d)). These illustrate that F adatoms close to the armchair and zigzag edges, respectively, create and destroy the same-spin arrangement there. The pristine ZGNRs, as indicated in Fig. 4(e), has an anti-ferromagnetic configuration in the absence of magnetic moment. Specifically, the coexistence of edge and center distributions is revealed in the central two-F adsorption (Fig. 4(f)), leading to an unusual anti-ferromagnetic configuration with a zero magnetic moment.
          
           \begin{figure}[htb]
                              \includegraphics[width=10cm, height=15cm]{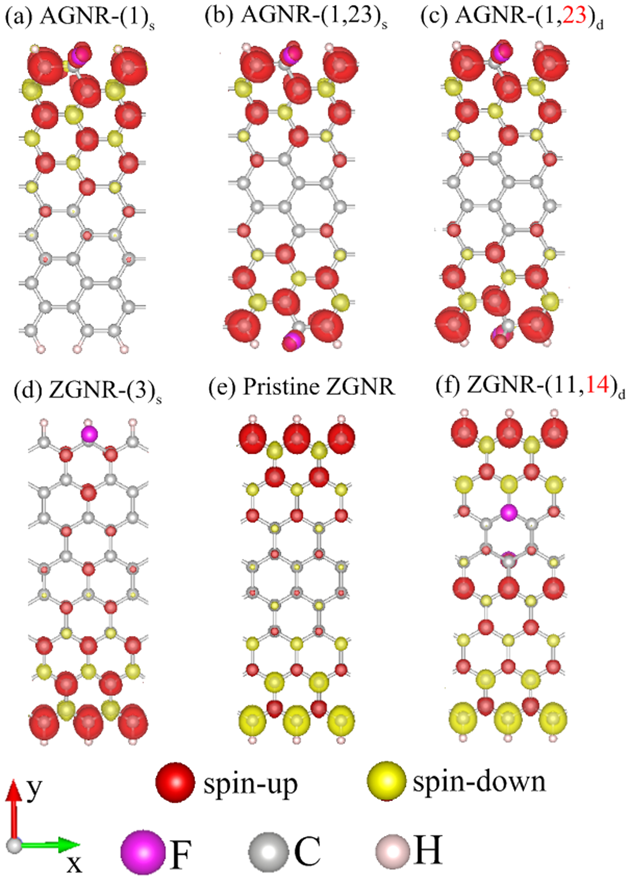}
                               \caption{Spin density of N\(_A=12\) AGNR for (a) \((1)_s\), (b) \((1, 23)_s\), (c) \((1,\textcolor{red}{23})_d\), and N\(_Z=8\) ZGNR for (d) \((3)_s\), (e) pristine, and (f) \((11,\textcolor{red}{14})_d\).}
                               \label{fgr:4}
                               \end{figure}

           In addition to spin arrangements, the charge density  \((\rho ) \)
             , the charge density difference 
             \((\Delta \rho ) \)
              , and the partial charge density 
              \(
              (\rho_P )
              \)
                are very useful in understanding the multi-orbital hybridizations in chemical bonds and the fluorination-enriched energy bands.  \(\Delta \rho \) is generated by subtracting the charge density of GNR and F adatoms from that of F-absorbed system. \(\rho \)  clearly illustrates the chemical bondings as well as the charge transfer. For a planar GNR, the parallel 2p\(_z\) orbitals and the planar (2p\(_x\),2p\(_y\),2s) orbitals, respectively, form the \(\pi\) and \(\sigma\) bondings, as shown in Fig. 5(a) for \(N_A=12\) AGNR (the solid and dashed rectangles). Fluorination can induce the high charge density between F and C, obvious change in \(\pi\) bonding, and observable reduction in \(\sigma\) bonding (Figs. 5(b) and 5(d)). The strong fluorination effects are also revealed in drastic density variations, especial for \((\Delta \rho ) \)  near F adatoms on 
                \((x,z)\) and \((y,z)\)
                  and   planes (Figs. 5(c) and 5(e)). These clearly indicate the complicated (2p\(_x\),2p\(_y\),2p\(_z\))-(2p\(_x\),2p\(_y\),2p\(_z\)) hybridizations in F-C bonds. When the distance between two fluorine adatoms is sufficiently short, there exist the significant (2p\(_x\),2p\(_y\)) hybridizations in F-F bonds, as illustrated in Figs. 5(c) and 5(e) (red rectangles on \((x,z)\) and \((y,z)\)
                  planes). As for the metallic and semiconducting behaviors, they are, respectively, characterized by the distorted and terminated \(\pi\) bondings in the partial charge density related to electronic states very close to \(E_F\)  (Figs. 5(f) \(\&\) 5(h), and Figs. 5(g) \(\&\) 5(i)).
                  
                    The diverse electronic structures and magnetic configurations are clearly evidenced in the orbital- and spin-projected DOSs (Fig. 6). There are a plenty of special structures, in which the asymmetric and symmetric peaks, respectively, come from the parabolic and partially flat bands. When the F-concentration is below 
                    \(
                    50\%\) (Figs. 6(a)-6(b) \(\&\) 6(e)-6(g)), the low-energy DOS is dominated by the \(\pi\) bonding of C-2p\(_z\) orbitals (red curves). This bonding also makes contributions to the deeper-energy DOS. The peak structures, which is due to the \(\sigma\) bonding of (2p\(_x\),2p\(_y\)) orbitals, appear at 
                     \(
                     E <  - 2.5\)
                      eV (green curves). The similar structures are revealed by the (2p\(_x\),2p\(_y\)) and 2p\(_z\) of the F adatoms (dashed blue and pink curves). The former present a sufficiently wide energy width of 2 eV, so that there exist the significant (2p\(_x\),2p\(_y\)) orbital hybridizations in F-F bonds. All the orbitals can create the merged peak structures at deeper energy, clearly illustrating the (2p\(_x\),2p\(_y\),2p\(_z\))-(2p\(_x\),2p\(_y\),2p\(_z\)) multi-orbital hybridizations in F-C bonds. Specifically, at high concentrations, the energy width of the bands might be more than \
                      \(5\) eV (Figs. 6(c)-6(d)); furthermore, the corresponding peaks are stronger than those of the \(\sigma\) bands. The occupied valence bands are closely related to the (2p\(_x\),2p\(_y\)) orbitals of F and C, especially for the orbital hybridization in F-F bonds.

                 \begin{figure}[htb]
                               \includegraphics[width=8cm, height=20cm]{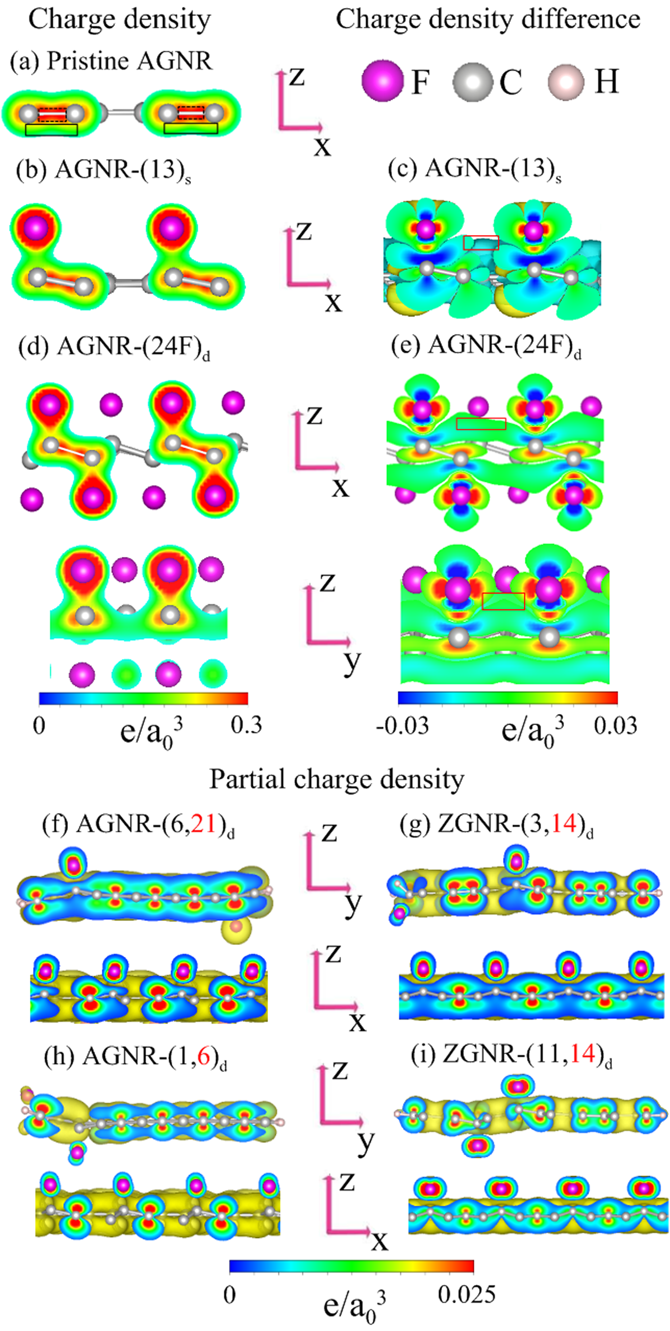}
                                 \caption{Spatial charge density of N\(_A=12\) AGNR for (a) Pristine, (b) \((13)_s\), \(\&\) (d) (24F)\(_d\); charge density difference for (c) \((13)_s\), (e) (24F)\(_d\). Partial charge density is shown for (f) AGNR-\((6,\textcolor{red}{21})_d\), (g) ZGNR-\((3,\textcolor{red}{14})_d\), (h) AGNR-\((1,\textcolor{red}{6})_d\), and (i) ZGNR-\((11,\textcolor{red}{14})_d\).}
                                     \label{fgr:5}
                              \end{figure}

               Five kinds of essential properties are characterized by the specific peak structures near the Fermi level. A pair of anti-symmetric peaks near \(E=0\), which is divergent in the opposite direction, is characteristic of energy gap in the absence of spin splitting (the first kind), as shown in Figs. 6(a)-6(d). But for non-magnetic metals (the second kind), the similar pair presents a blue shift about \(0.5 - 1.0\) 
                eV (Figs. 6(e) and 6(f), and DOS is finite at the Fermi level. The spin-polarized peak structures are revealed in ferromagnetic metals (the third kind in Figs. 6(g) and 6(h)). The low-energy peaks are quite different for spin-up and spin-down configurations, in which they are asymmetric about \(E=0\), and the former predominates the occupied states of \(E<0\). Specifically, the partially flat bands in a pristine ZGNR can create a pair of symmetric peaks (blue triangles in Fig. 6(i)), accompanied with an energy gap and one peak due to an extra band-edge state (Fig. 3(a)). This corresponds to an anti-ferromagnetic semiconductor with spin degeneracy (the fourth kind). Three are more pairs of symmetric peaks centered about   in the presence of spin splitting, as indicated in Fig. 6(j). The fifth kind of peak structure arises from narrow-gap F-absorbed ZGNRs with the spin-split anti-ferromagnetism.
               
                 \begin{figure}[htb]
                                 \includegraphics[width=10cm, height=20cm]{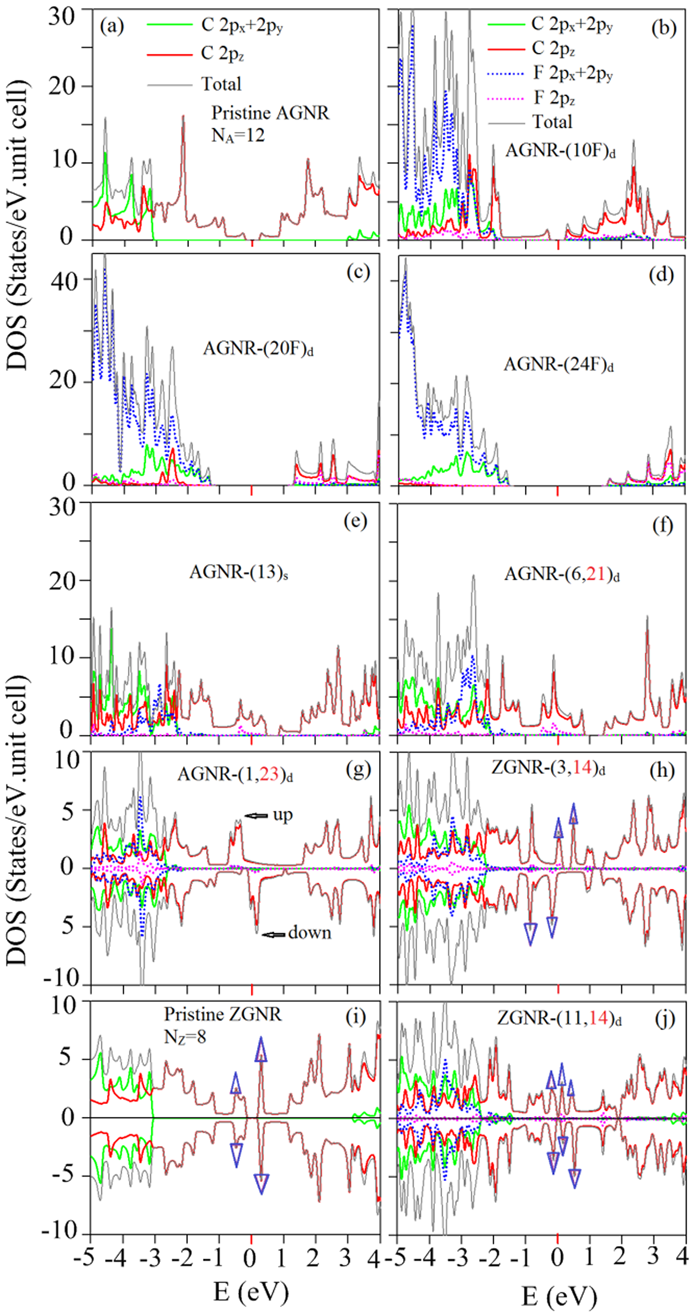}
                                  \caption{Orbital-projected DOSs for (a) Pristine, (b) (10F)\(_d\)-, (c) (20F)\(_d\)-, (d) (24F)\(_d\)-, (e) \((13)_s\)-, (f) \((6,\textcolor{red}{21})_d\)-, (g) \((1,\textcolor{red}{23})_d\)-adsorbed N\(_A=12\) AGNR;  (h) \((3,\textcolor{red}{14})_d\)-, (i) pristine;  (j) \((11,\textcolor{red}{14})_d\)-adsorbed N\(_Z=8\) ZGNR. Blue triangles correspond to the partially flat bands.}
                                   \label{fgr:6}
                                  \end{figure}

               On the experimental sides, STM, which can provide the spatially atomic distributions of the local nano-structures, have been successfully used to identify the unique geometric structures of the graphene-related systems, covering graphite, graphene, graphene compounds, carbon nanotubes, and GNRs.  The atomic-scaled observations clearly reveal the 2D networks of local defects \cite{48}, the buckled and rippled structures of graphene islands \cite{49,50,51}, the adatom distributions on graphene surface \cite{52}, the nanoscale width of GNR \cite{53}, and the chiral arrangements of the hexagons on the planar edges \cite{54} and a cylindrical surface \cite{55}. As to F-absorbed GNRs, the buckled structure, the adatom height, and the bond length (Table 1) deserve further STM examinations. Such measurements are very useful in the identification of the significant orbital hybridizations related to F-C, F-F and C-C chemical bonds.
                        
               ARPES is the most powerful experimental technique to examine the wave-vector-dependent band structures. The experimental measurements on graphene-related systems have confirmed the feature-rich electronic structures under the distinct dimensions. For example, the verified energy bands include an isotropic Dirac-cone structure with linear energy dispersions in monolayer graphene \cite{56}, two pairs of parabolic bands in bilayer AB stacking \cite{57}, the bilayer- and monolayer-like energy dispersions, respectively, at \(k_z=0\) and zone boundary in AB-stacked graphite \cite{58}, and 1D parabolic energy bands with energy gaps in AGNRs \cite{59}. In addition, an edge-localized partial flat band is deduced to be associated with the zigzag-like steps on graphite surface \cite{60}. Up to now, the ARPES measurements on the adatom-enriched unusual energy bands of GNRs are absent. The ARPES and spin-resolved ARPES \cite{61} are available in verifying the predicted five kinds of electronic structures and magnetic configurations in F-absorbed GNRs. That is to say, the complicated relations among the finite-size effects, the edge structures, the spin configurations, and the multi-orbital hybridizations could be examined by them.
               
                The STS measurements, with the tunneling differential conductance 
                \((dI/dV)\) proportional to DOS, could serve as very efficient methods to identify the dimension-enriched special structures in DOS. They have verified the diverse electronic properties in sp\(^2\)-bonding carbon-related systems.  The measured DOSs show the splitting \(\pi\) and 
                \(
                \pi ^ *\)  
                 peaks and a finite value near \(E_F\) characteristic of the semi-metallic behavior in graphite \cite{62}, a linear E-dependence vanishing at the Dirac point in monolayer graphene, the asymmetry-created peak structures in bilayer graphene \cite{63},  a prominent peak at \(E_F\) arising from partially flat bands in tri-layer and penta-layer ABC-stacked graphene \cite{64}; the geometry-dependent energy gaps and the asymmetric peaks of 1D parabolic bands in carbon nanotubes \cite{65} and GNRs. The STS and spin-resolved STS could be utilized to examine five kinds of low-lying DOSs in F-absorbed GNRs, covering the finite value at \(E_F\), energy gap and spin-polarized peak structures. In short, geometric structures, electronic properties and magnetic configurations are enriched by the multi-orbital hybridizations in strong chemical bondings and the spin arrangements. The interactions of atomic orbitals present 2p\(_z\)-2p\(_z\) and (2p\(_x\),2p\(_y\))-(2p\(_x\),2p\(_y\)) in C-C bonds, (2p\(_x\),2p\(_y\),2p\(_z\))-(2p\(_x\),2p\(_y\),2p\(_z\)) in F-C bonds, and  (2p\(_x\),2p\(_y\))-(2p\(_x\),2p\(_y\)) in F-F bonds.

        \section{Concluding remarks}
          \par\noindent
         
          The geometric structures, electronic and magnetic properties of F-adsorbed GNRs are investigated using the first-principles calculations. The atom-dominated band structure, the spatial charge density, the spin arrangement, and the orbital-projected DOS are useful in exploring the orbital- and spin-dependent essential properties. For example, band structure, free carrier density, magnetism and DOS are determined by which kinds of orbital hybridizations and spin configurations. The similar analyses could be further generalized to the emergent layered materials, with nanoscale thickness and unique lattice symmetries, covering silicene, germanene, tinene, phosphorene, MoS\(_2\) and so on. The fluorination-induced diverse phenomena mainly arises from the complicated relations among lattice symmetry, quantum confinement, edge structure, significant chemical bonding, and spin arrangement. The metallic or semiconducting behaviors with/without magnetism indicate the highly potential applications, such as electronic, optical, and spintronic devices.
           
           Each F-absorbed GNR presents an obvious buckling, in which the adatom height and the change of C-C bond length are mainly determined by the very strong fluorination. The multi-orbital hybridizations in F-C, C-C and F-F bonds, and the edge-dependent spin distributions are responsible for five kinds of electronic and magnetic properties. The metallic behavior is revealed at certain F-distributions below 
           \(40\% \)
            concentration. AGNRs could create one or two holes per unit cell, respectively, corresponding to the spin-split and spin-degenerate \(\pi\)-electronic energy bands (non-magnetism and ferromagnetism). However, ZGNRs only exhibit the lower carrier-density case. On the other hand, the semiconducting systems could survive at various F-adsorptions except for single adatom. The \(\pi\) bonding predominates the essential properties, including the gap-dependent parabolic bands with spin degeneracy (non-magnetism) and partially flat bands in the absence/presence of spin splitting (anti-ferromagnetism). Under high adatom concentrations, non-magnetic parabolic bands, with an energy gap are determined by the (2p\(_x\),2p\(_y\)) orbitals of F and C. The 1D energy dispersions are reflected in DOS as many anti-symmetric and symmetric peaks. There exist five kinds of low-lying peak structures characteristic of the main features of essential properties. The predicted geometric structures, energy bands and DOSs, could be verified by STM, ARPES and STS, respectively, especially for the latter two tools with spin resolution.
            
            \newpage
            \par\noindent {\bf Acknowledgments}
            
            This work was supported by the Physics Division, National Center for Theoretical Sciences (South), the Nation Science Council of Taiwan. We also thank the National Center for High-performance Computing (NCHC) for computer facilities.
            \newpage
            \renewcommand{\baselinestretch}{0.2}

\end{document}